\documentstyle[12pt]{article}

\newcommand{\vect}[1]{\stackrel{\rightarrow}{#1}}
\newcommand{\bi}{\bibitem}
\font\ab= msbm10 scaled\magstep1
\begin{document}
\title{Schr\" odinger  quantum modes  on the Taub-NUT background}

\author{Ion I. Cot\u aescu \thanks{E-mail:~~~cota@quasar.uvt.ro}\\ 
{\small \it The West University of Timi\c soara,}\\
       {\small \it V. P\^ arvan Ave. 4, RO-1900 Timi\c soara, Romania}
\and
Mihai Visinescu \thanks{E-mail:~~~mvisin@theor1.theory.nipne.ro}\\
{\small \it Department of Theoretical Physics,}\\
{\small \it National Institute for Physics and Nuclear Engineering,}\\
{\small \it P.O.Box M.G.-6, Magurele, Bucharest, Romania}}
\date{\today}

\maketitle

\begin{abstract}

The Schr\" odinger equation is investigated in the Euclidean Taub-NUT 
geometry. The bound states are degenerate and an extra degeneracy is 
due to the conserved Runge-Lenz vector. The existence of the extra 
conserved quantities, quadratic in four-velocities implies the 
possibility of separating variables in two different coordinate 
systems. The eigenvalues and the eigenvectors are given in both cases 
in explicit, closed form.

Pacs 04.62.+v

\end{abstract}
\

\section{Introduction}
\
The discovery of the self-dual instanton solutions to Euclidean 
Yang-Mills theory suggested the possibility that analogous solutions to 
the Euclidean Einstein equations might be important in quantum gravity.
One example of a metric which satisfies the Euclidean Einstein 
equations with self-dual Riemann tensor is the self-dual Taub-NUT 
metric \cite{Ha}. In this case Einstein's equations are satisfied with 
zero cosmological constant and the manifold is {\ab R}$^4$ with a 
boundary which is a twisted three-sphere $S^3$ possessing a distorted 
metric.

On the other hand much attention has been paid to the Euclidean Taub-NUT 
metric because it is involved in many modern studies in physics. The 
Kaluza-Klein monopole of Gross and Perry \cite{GP} and of Sorkin 
\cite{So} was obtained by embedding the Taub-NUT gravitational instanton 
into five-dimensional Kaluza-Klein theory. Remarkably, the same object 
has re-emerged in the study of monopole scattering. In the 
long-distance limit, neglecting radiation, the relative motion of two 
monopoles is described by the geodesics of this space \cite{Ma, AH}. 

It is well known that space-time isometries give rise to constants of 
motion along geodesics. However it is worth to mention that not all 
conserved quantities arose from isometries of the manifolds and 
associated Killing vector fields. Other integrals of motion are related 
to "hidden" symmetries of the manifold, which manifest themselves as 
tensor of rank $n>1$, satisfying a generalized Killing equation, namely 
$\nabla_{(\mu} K_{\nu_1 ...\nu_n)} = 0$ \cite{GH}. They are usually 
referred to as St\" ackel-Killing tensors.

An illustration of the existence of extra conserved quantities is 
provided by the Taub-NUT geometry. Indeed, for the geodesic motions in 
the Taub-NUT space, there is a conserved vector, analogous to the 
Runge-Lenz vector of the Kepler type problem, whose existence is rather 
surprising in view of the complexity of the equations of motion  
\cite{GM, GR, FH, CFH}. The Runge-Lenz vector is quadratic in 
four-velocities and its components are St\" ackel-Killing tensors.

The Taub-NUT space is also of mathematical interest. In the Taub-NUT 
geometry there are known to exist four Killing-Yano tensors. The 
Killing-Yano tensor \cite{Ya} here is a $2$-form, $f_{\mu\nu} 
= f_{[\mu\nu]}$, which satisfies the Penrose-Floyd equation 
$\nabla_{(\mu} f_{\nu)\lambda} = 0$. Three of these Killing-Yano tensors 
are rather special: they are covariantly constant, mutually 
anti-commuting and square the minus unity. Thus they are complex 
structure realizing the quaternionic algebra and the Taub-NUT 
manifold is hyper-K\" ahler \cite{GR}. The fourth 
Killing-Yano tensor has a non-vanishing field strength, is not trivial 
and leads to new constants of motion. More precisely, the symmetrized 
product of the fourth Killing-Yano tensor with the previous three is 
connected with the St\" ackel-Killing tensors, i.e. the components of 
the Runge-Lenz vector \cite{GR, vH, VV1,VV2}.

The aim of this paper is to study the Schr\" odinger equation in 
Taub-NUT space. The conserved Runge-Lenz vector implies an extra 
degeneracy of the bound states. It is remarkably the fact that there
appears to be a close relation between the existence of extra conserved 
quantities and the possibility of separating variables in two different 
coordinate systems. In this way one obtains two different orthonormal 
energy bases, called here the central and axial bases, for which the 
degeneracy of the energy levels  is  well defined by the eigenvalues of 
suitable sets of commuting operators.    

The plan of the paper is as follows. In Section 2 we summarize the 
relevant features of the Euclidean Taub-NUT geometry. In Section 3 we 
analyze the Schr\" odinger equation in this space looking for the 
complete  sets of commuting operators that define the mentioned bases. 
In the next two sections 
the Schr\" odinger equation is investigated in spherical and parabolic 
coordinates. We show that the central modes can be solved in spherical 
coordinates while for the axial modes we need to use parabolic coordinates. 
In both cases we obtain sets of regular modes as well as specific sets of 
irregular modes arising from the special boundary conditions of the 
Taub-NUT geometry.  Our concluding remarks are presented in the last section. 
Finally, the Appendix is devoted to $SO(3)\otimes U(1)$ harmonics.

\section{Taub-NUT geometry}
\

The Euclidean Taub-NUT manifold is a particular case of static Euclidean 
Kaluza-Klein space-time  whose metric, in a static chart of  coordinates, 
$t$ and $x^{\mu}$, ($\mu, \nu,...=1,2,3,5$), is given by the line element
\begin{equation}\label{(met)} 
ds^{2}=-dt^{2}+g_{\mu\nu}dx^{\mu}dx^{\nu}=-dt^{2}+\frac{1}{V}dl^{2}+V(dx^{5}+
A_{i}dx^{i})^{2}\,,
\end{equation}   
where 
\begin{equation}
dl^{2}=(d\vect{x})^{2}=(dx^{1})^{2}+(dx^{2})^{2}+(dx^{3})^{2}
\end{equation}
is the usual Euclidean 3-dimensional line element involving the Cartesian 
space coordinates $x^{i}$ ($i,j,...=1,2,3$) which 
cover the domain $D$. The other coordinates are the time, $t$, and 
the Cartesian Kaluza-Klein extra-coordinate, $x^{5}\in D_{5}$. We suppose that 
$V$ and $A_{i}$ are static functions depending only on $\vect{x}$. It is clear 
that in the absence of the scalar potential we have  
\begin{equation}
{\rm div}\vect{A}\,=0\,, \quad {\rm rot}\vect{A}\,=\,\vect{B}.
\end{equation}  
>From (\ref{(met)}) we obtain the covariant components of the metric tensor 
\begin{equation}
g_{ij}=\frac{1}{V}\delta_{ij}+VA_{i}A_{j}\,,\quad 
g_{i5}=VA_{i}\,,\quad g_{55}=V\,,
\end{equation}
and
\begin{equation}
g={\rm det}(g_{\mu\nu})=\frac{1}{V^{2}}\,.
\end{equation}
The corresponding contravariant components  are
\begin{equation}
g^{ij}=V\delta_{ij}\,,\quad g^{i5}=-VA_{i}\,,\quad 
g^{55}=\frac{1}{V}+V{\vect{A}\,}^{2}\,.
\end{equation}
   
The Euclidean Taub-NUT manifold has the line 
element of the form (\ref{(met)}) with
\begin{equation}\label{(tn)}
V^{-1}=1+\frac{\mu}{r}\,,\quad A_{1}=-\frac{\mu}{r}\frac{x^{2}}{r+x^{3}}\,,
\quad A_{2}=\frac{\mu}{r}\frac{x^{1}}{r+x^{3}}\,,\quad A_{3}=0\,, 
\end{equation}
where $r=|\vect{x}|$ and $\mu$ is a real number. Hereby, it results  the 
magnetic field with central symmetry
\begin{equation}
\vect{B}\,=\mu\frac{\vect{x}}{r^3}\,.
\end{equation}
In fact, the geometry defined by (\ref{(met)}) and (\ref{(tn)}) has the 
{\em global symmetry} 
of the group $SO(3)\otimes U_{5}(1)\otimes T_{t}(1)$. This means that  
the line element is invariant under global rotations
of the Cartesian space coordinates  
and  $x^{5}$ and $t$ translations of the Abelian groups 
$U_{5}(1)$ and $T_{t}(1)$ respectively. 
The generators of these groups are just the  five Killing vectors 
corresponding to this global symmetry. Moreover, there is a three-component 
Killing tensor $\vect{k}_{\mu\nu}$ which satisfies $\nabla_{(\sigma}
\vect{k}_{\mu\nu)}=0$ (where $\nabla$ is the covariant derivative and 
$~_{(~~)}$ denotes the symmetrisation).      
\newpage
\section{Observables}
\

The quantum mechanics  in the Taub-NUT geometry \cite{CFH} is based on the 
Schr\" odinger equation   
\begin{equation}\label{(kg)}
H\psi=i\partial_{t}\psi
\end{equation}
involving the Hamiltonian operator    
\begin{equation}
H=-\frac{1}{2}\nabla_{\mu}g^{\mu\nu}\nabla_{\nu}\,,
\end{equation}
written in natural units with $\hbar=c=1$.
In any static chart,  Eq.(\ref{(kg)}) has  the particular solutions   
\begin{equation}\label{(sol)}
\psi_{E}(x)=U_{E}(\vect{x},x^{5})e^{-iEt}\,, 
\end{equation}
where $U_{E}$ are energy eigenfunctions,  
\begin{equation}\label{(he)}
HU_{E}=EU_{E}\,.
\end{equation}
The solutions (\ref{(sol)}) may be either square integrable 
functions or tempered distributions on  $D\times D_{5}$. In  both cases they 
must be orthonormal (in usual or generalized sense) with respect to the 
scalar product  
\begin{equation}\label{(scpr)} 
\left<\psi,\psi'\right>
=\int_{D\times D_{5}}d^{4}x\,\sqrt{g}\,
\psi^{*} \psi'\,.
\end{equation}
We denote by ${\cal L}^{2}$ the Hilbert space  of the square integrable 
functions with respect to this scalar product.

The main operators on  ${\cal L}^{2}$  can be introduced by using the 
geometric quantization \cite{CFH}. In this way, one obtains  the 
non-hermitian momentum operators in the coordinate representation, 
\begin{equation}
P_{i}=-i(\partial_{i}-A_{i}\partial_{5})\,,\quad P_{5}=-i\partial_{5}\,,
\end{equation} 
which obey the commutation rules
\begin{equation}
[P_{i},P_{j}]=i\epsilon_{ijk}B_{k}P_{5}\,,\quad
[P_{i},P_{5}]=0\,.
\end{equation}
With their help we can write  
\begin{equation}
H=\frac{1}{2}\left(V{\vect{P}\,}^{2}+\frac{1}{V}{P_{5}}^{2}\right)\,.
\end{equation}
Hereby we see that the $U_{5}(1)$ generator, $P_{5}$, is conserved since it 
commutes with $H$. This is natural since $P_{5}$ is up to the factor $-i$ 
just a Killing vector.   In the following it is convenient to replace it by 
$Q=-\mu P_{5}$. Furthermore, one can verify that other three Killing 
vectors are the components of the angular momentum operator 
\begin{equation}\label{(angmom)}
\vect{L}\,=\,\vect{x}\times\vect{P}+\frac{\vect{x}}{r}Q\,.
\end{equation} 
These  are  conserved  and satisfy the canonical commutation rules. 
Moreover, their commutators with the coordinates and the momentum operators 
are the usual ones. Other three important conserved operators are those 
defined by the three-component Killing tensor $\vect{k}_{\mu\nu}$ as 
\cite{C,CFH} 
\begin{equation}
\vect{K}\,=-\frac{1}{2}\nabla_{\mu}\vect{k}^{\mu\nu}\nabla_{\nu}=
\frac{1}{2}\left(\vect{P}\times \vect{L}-
\vect{L}\times \vect{P}\right)-
\mu \frac{\vect{x}}{r}
\left(H-\frac{1}{\mu^{2}}Q^{2}\right)\,.
\end{equation}
Thus we obtain a vector operator which play the same role as the Runge-Lenz 
vector in the usual quantum mechanical Kepler problem. This transforms as a 
vector under space rotations since its components obey
\begin{equation}
[L_{i},K_{j}]=i\epsilon_{ijk}K_{k}\,.
\end{equation}

Now we have to chose the suitable complete sets of commuting operators which 
should define usual or generalized bases for the Hilbert space ${\cal L}^{2}$. 
Here two options are interesting, namely (I) that of the {\em central} basis 
given by the set $\{H,{\vect{L}\,}^{2},L_{3},Q\}$ or (II) the {\em axial} 
basis formed by the common eigenfunctions of the set $\{  
H,K_{3},L_{3},Q\}$. In both cases we must work in suitable coordinate  systems
which should allow us to separate the variables in all the differential 
equations corresponding to the eigenvalue problems.
\newpage
\section{Central modes} 
\

In order to study the central modes it is convenient to choose the local 
chart with spherical coordinates,  $r, \theta, \phi$, commonly related 
to the Cartesian ones  and the new coordinate $\chi$ defined as 
\begin{equation}
\chi=-\frac{1}{\mu}x^{5}-{\rm arctan}\frac{x^2}{x^1}\,.
\end{equation}
Here the Taub-NUT  line element is 
\begin{equation}
ds^{2}=-dt^{2}+\frac{1}{V}(dr^{2}+r^{2}d\theta^{2}+ 
r^{2}\sin^{2}\theta\, d\phi^{2})+\mu^{2}V(d\chi+\cos\theta\, d\phi)^{2}\,,  
\end{equation}
since
\begin{equation}
A_{r}=A_{\theta}=0\,,\quad A_{\phi}=\mu(1-\cos\theta)\,.
\end{equation}
Moreover, we consider that in this chart  $r\in D_{r}=\{r|V(r)>0\}$ (i.e., 
$r>0$ if $\mu>0$ or $r>|\mu|$ if $\mu<0$), the angular coordinates 
$\theta,\,\phi$ cover the sphere $S^{2}$ while $\chi\in D_{\chi}=[0,4\pi)$. 

The main observables are 
\begin{equation}
Q=-i\partial_{\chi}
\end{equation}
and  the components of the angular momentum (\ref{(angmom)}) in 
the canonical  basis (with $L_{\pm}=L_{1}\pm iL_{2}$),
\begin{eqnarray}
L_{3}&=&-i\partial_{\phi}\,,\\
L_{\pm}&=&e^{\pm i\phi}\left[\pm\,\partial_{\theta}+
i\left(\cot\theta\,\partial_{\phi}-
\frac{1}{\sin\theta}\,\partial_{\chi}\right)\right]\,.
\end{eqnarray}
The $SO(3)$ Casimir operator 
\begin{eqnarray}
{\vect{L}\,}^{2}&=& L_{+}L_{-}+L_{3}^{2}-L_{3}\label{(lpa)}\\
&=&-\frac{1}{\sin\theta}\,\partial_{\theta}\,(\sin\theta\,\partial_{\theta})-
\frac{1}{\sin^{2}\theta} (\partial_{\phi}^{2}+\partial_{\chi}^{2}-
2\cos\theta\,\partial_{\phi}\,\partial_{\chi})\nonumber
\end{eqnarray}
is related to the momentum operators through
\begin{equation}
-{\vect{P}\,}^{2}={\partial_{r}}^{2}+\frac{2}{r}\partial_{r}-
\frac{1}{r^2}{\vect{L}\,}^{2}+\frac{1}{r^2}Q^{2}\,.
\end{equation}

Looking for the common eigenfunctions of the set (I),
we observe that  the angular momentum as well as $Q$ act only upon 
functions of $\theta,\,\phi$ and $\chi$ which must be square integrable since 
the domain  $S^{2}\times D_{\chi}$ is compact. These form the  Hilbert space  
${\cal L}^{2}(S^{2}\times D_{\chi})$ where the complete set of commuting 
operators $\{{\vect{L}\,}^{2},L_{3},Q\}$ determines a suitable basis. Their 
common eigenfunctions, $Y^{q}_{l,m}$, which satisfy the eigenvalue problems
\begin{eqnarray}
{\vect{L}\,}^{2}Y_{l,m}^{q}&=&l(l+1)\,Y_{l,m}^{q}\,,\label{(lp)}\\
L_{3}Y_{l,m}^{q}&=&m\,Y_{l,m}^{q}\,,\label{(l3)}\\
QY_{l,m}^{q}&=&q\,Y_{l,m}^{q}\,,\label{(l0)}
\end{eqnarray}
and the orthonormalization condition 
\begin{eqnarray}
&&\left<Y_{l,m}^{q},Y_{l',m'}^{q'}\right>=
\int_{S^2}d(\cos\theta)d\phi\,\int_{0}^{4\pi}d\chi\,
{Y_{l,m}^{q}(\theta, \phi, \chi)}^{*}\,
Y_{l',m'}^{q'}(\theta, \phi, \chi)\nonumber\\
&&~~~~~~~~~~~~~~~~~~~~~~~~~~~~~=\delta_{l,l'}\delta_{m,m'}
\delta_{q,q'}\,,\label{(spy)}
\end{eqnarray}
will be called  $SO(3)\otimes U(1)$ harmonics.   
Notice that the boundary conditions on $S^{2}\times D_{\chi}$ require $l$ and 
$m$ to be integer numbers while $q=0,\pm 1/2,\pm 1,...$ \cite{CFH}. The form 
of these harmonics is given in Appendix.

Thus the problem of the angular motion is completely solved and it is clear 
that the angular coordinates can be separated from the radial one if 
we take the common eigenfunctions of the set (I) of the form
\begin{equation}
U^{q}_{E,l,m}(r,\theta,\phi,\chi)=\frac{1}{r}R^{q}_{E,l}(r)Y^{q}_{l,m}
(\theta,\phi,\chi)\,.
\end{equation}
Then from Eq.(\ref{(he)}) we obtain the familiar radial equation,
\begin{equation}\label{(rad)}
\left[-\frac{d^2}{dr^2}+\frac{l(l+1)}{r^2}-\frac{\alpha}{r}\right]
R_{l,m}^{q}(r)=\beta R_{l,m}^{q}(r)
\end{equation}
where we have denoted
\begin{equation}\label{(ab)}
\alpha=2\mu\left[E-\frac{q^2}{\mu^2}\right]\,,\quad
\beta=2E-\frac{q^2}{\mu^2}\,.
\end{equation}
The radial wave functions  must be orthonormal with 
respect to the radial scalar product 
\begin{equation}\label{(scprod1)}
\left<R_{E,l}^{q},R_{E',l}^{q}\right>=\int_{D_{r}}dr \,\left(1+\frac{\mu}{r}
\right) R_{E,l}^{q}(r)^{*} R_{E',l}^{q}(r)\,.
\end{equation} 

This is a  Kepler-like problem similar to that of the 
non-relativistic quantum mechanics. The general solution 
of Eq.(\ref{(rad)}) can be written in terms of the confluent hypergeometric 
function as   
\begin{equation}\label{(rada)}
R_{E,l}^{q}(r)=N \rho^{s}e^{-\rho}F(a,2s,\rho)\,,
\end{equation}
where
\begin{equation}
\rho=2r\sqrt{-\beta}\,,\quad a=s-\frac{\alpha}{2\sqrt{-\beta}}\,,
\end{equation}
and $N$ is the normalization constant. The parameter $s$ is a solution of 
the equation $s(s-1)=l(l+1)$. The modes with $s=l+1$ are similar to those of 
the usual non-relativistic case. 
For this reason we say that these are {\em regular} modes. We shall see that  
in the Taub-NUT geometry there is a conjecture in which {\em irregular} modes 
with $s=-l$ are also allowed. In the following we shall briefly discuss the 
corresponding energy spectra assuming that $E\ge 0$.
  
Let us consider first the regular modes ($s=l+1$). It is easy to   
show that for $\mu>0$ there is only a continuous energy spectrum, covering 
the domain $E\ge q^{2}/2\mu^{2}$, the levels of which are infinite degenerated 
as it result from the selection rules (\ref{(qml)}). Notice that these rules 
are in accordance with the behavior of the classical trajectories which 
are open and cannot reach the center.   
In the case of $\mu<0$ we have the same continuous energy spectrum with 
infinite degenerate levels and, in addition, a discrete spectrum in 
the domain $[0,\, q^{2}/2\mu^{2})$ where $\alpha>0$ and $\beta<0$. Here we can 
impose the quantization condition $a=-n_{r}, \,n_{r}=0,1,2,...$ which gives  
\begin{equation}\label{(abn)}
\frac{\alpha}{\sqrt{-\beta}}=2n
\end{equation} 
where $n=n_{r}+l+1$ is the main quantum number of the regular modes. 
Hereby we find the energy levels 
\begin{equation}\label{(een)}
E_{n}=\frac{1}{\mu^2}\left[n\sqrt{n^{2}-q^{2}}-(n^{2}-q^{2})\right]
\end{equation}        
for all $n> |q|>0$ \cite{CFH}. These levels are finite degenerated since for 
given 
$q$ and $n$  the angular quantum numbers can take all the values selected by 
the conditions $|q|-1<l\le n-1$ and (\ref{(qml)}). Thus one obtains 
a countable energy spectrum with the property 
\begin{equation}
\lim_{n\to \infty} E_{n}=\frac{q^2}{2\mu^2}\,.  
\end{equation}

The irregular modes arise only as discrete modes when $\mu<0$ since the 
radial functions (\ref{(rada)}) remain square integrable on 
$D_{r}=[|\mu|,\infty)$ for $s=-l$ if $a=-n_{r}$ and $|q|+l<n_{r}<2l$.   
The main quantum number of these modes, $n=n_{r}-l$, obeys  $|q|<n<l$ and 
gives the same quantization rule (\ref{(abn)}) and energy levels 
(\ref{(een)}). 
Therefore, the discrete spectrum of irregular modes coincides 
to that of the regular ones, spanning just the domain of energies where the 
classical motion has closed trajectories. The difference is that in the case 
of irregular modes the energy levels are infinite degenerated 
(since $l>n$) and there is no continuous energy spectrum.

\section{Axial modes}
\

Let us consider now the axial modes given by the eigenvalue problem of the 
set (II). This can be solved in a local chart with  parabolic  
coordinates, $\xi\in D_{\xi}$, $\eta\in D_{\eta}$ and $\phi\in [0,2\pi)$, 
defined by 
\begin{equation}
x^{1}=\sqrt{\xi\eta}\cos\phi\,,\quad
x^{2}=\sqrt{\xi\eta}\sin\phi\,,\quad
x^{3}=\frac{1}{2}(\xi-\eta)\,,
\end{equation}
such that 
\begin{equation}
r=\frac{1}{2}(\xi+\eta)\,.
\end{equation}
Here the domains $D_{\xi}$ and $D_{\eta}$ corresponding to $D_{r}$ are either 
$\xi,\eta>0$ for $\mu>0$ or $\xi,\eta>|\mu|$ if $\mu<0$. 
The line element reads 
\begin{equation}\label{(metpar)}
ds^{2}=-dt^{2}+\frac{1}{V}dl^{2}+\mu^{2}V
\left(d\chi+\frac{\xi-\eta}{\xi+\eta}\, d\phi\right)^{2}
\end{equation}
where
\begin{equation}
dl^{2}=\frac{\xi+\eta}{4\xi}d\xi^{2}
+\frac{\xi+\eta}{4\eta}d\eta^{2}+\xi\eta d\phi^{2}\,.
\end{equation}

Following the same procedure like in the case of the central modes we 
calculate how look the operators of the set (II) in these coordinates. 
It is not difficult to show that 
\begin{equation}\label{(hxy)}
H=\frac{1}{2\mu^2}Q^{2}-\frac{2}{\xi+\eta+2\mu}(X+Y)
\end{equation}
and
\begin{equation}\label{(kxy)}
K_{3}=\frac{2}{\xi+\eta+2\mu}\left[(\eta+\mu)X-(\xi+\mu)Y\right]
\end{equation}
where we have denoted
\begin{eqnarray}
X&=&\partial_{\xi}(\xi\partial_{\xi})-\frac{1}{4\xi}(L_{3}+Q)^{2}
-\frac{1}{4\mu}Q^{2}\,, \\
Y&=&\partial_{\eta}(\eta\partial_{\eta})-\frac{1}{4\eta}(L_{3}-Q)^{2}
-\frac{1}{4\mu}Q^{2}\,. 
\end{eqnarray}

The whole eigenvalue problem of  axial modes is
\begin{eqnarray}
HU^{q}_{E,\kappa,m}&=&EU^{q}_{E,\kappa,m}\,,\label{(21)}\\
K_{3}U^{q}_{E,\kappa,m}&=&\kappa U^{q}_{E,\kappa,m}\,,\label{(22)}\\
L_{3}U^{q}_{E,\kappa,m}&=&mU^{q}_{E,\kappa,m}\,,\\
QU^{q}_{E,\kappa,m}&=&qU^{q}_{E,\kappa,m}\,.
\end{eqnarray}
These equations can be solved in parabolic coordinates if we put 
\begin{equation}
U^{q}_{E,\kappa,m}(\xi,\eta,\phi,\chi)=N f^{q}_{E,\kappa,m}(\xi)
h^{q}_{E,\kappa,m}(\eta)\,
e^{im\phi}\,e^{iq\chi}\,,
\end{equation}
where $N$ is the normalization constant  calculated with the help of the 
scalar product (\ref{(scpr)}) rewritten in parabolic coordinates as
\begin{eqnarray}
\left<U,U'\right>&=&\frac{1}{4}\int_{D_{\xi}}d\xi\int_{D_{\eta}}d\eta 
(\xi+\eta+2\mu)\nonumber\\
&&\times \int_{0}^{2\pi}d\phi\int_{0}^{4\pi}d\chi\,
U^{*}(\xi,\eta,\phi,\chi) 
U'(\xi,\eta,\phi,\chi)\,. 
\end{eqnarray}
By taking into account that the operators of Eqs.(\ref{(21)}) and (\ref{(22)}) 
have the form (\ref{(hxy)}) and (\ref{(kxy)}) respectively, we find that $f$ 
and $h$ satisfy
\begin{eqnarray}
\left[\partial_{\xi}(\xi\partial_{\xi})-\frac{(m+q)^2}{4\xi}+
\frac{\alpha}{4}+\frac{\beta}{4}\xi\right]f^{q}_{E,\kappa,m}(\xi)&=&
\frac{\kappa}{2}\,f^{q}_{E,\kappa,m}(\xi)\,,\\
\left[\partial_{\eta}(\eta\partial_{\eta})-\frac{(m-q)^2}{4\eta}+
\frac{\alpha}{4}+\frac{\beta}{4}\eta\right]h^{q}_{E,\kappa,m}(\eta)&=&
-\frac{\kappa}{2}\,h^{q}_{E,\kappa,m}(\eta)\,,
\end{eqnarray}
where $\alpha$ and $\beta$ are the functions of $E$ and $q$ defined by 
Eqs.(\ref{(ab)}). 
The last step is to solve these equations. Defining the new variables 
\begin{equation}
x=\xi\sqrt{-\beta}\,,\quad y=\eta\sqrt{-\beta}\,,
\end{equation}
we obtain the general form of the solutions,  
\begin{eqnarray}\label{(fh)}
f^{q}_{E,\kappa,m}(\xi)&=&\,x^{s_{1}}\,e^{-x/2}F\left(a_{1}, 2s_{1}+1,
\,x\right)\,,\\
h^{q}_{E,\kappa,m}(\eta)&=&\,y^{s_{2}}\,e^{-y/2}F\left(a_{2}, 
2s_{2}+1,\,y\right)\,,
\end{eqnarray}
where
\begin{eqnarray}
a_{1}&=&s_{1}+\frac{1}{2}+\frac{2\kappa-\alpha}{4\sqrt{-\beta}}\,,\\ 
a_{2}&=&s_{2}+\frac{1}{2}-\frac{2\kappa+\alpha}{4\sqrt{-\beta}}\,, 
\end{eqnarray}
depend on the parameters $s_{1}=\pm (|m|+q)/2$ and $s_{2}=\pm(|m|-q)/2$. 

We define the regular axial modes for $s_{1}=(|m|+q)/2>-1/2$ and 
$s_{2}=(|m|-q)/2>-1/2$ which means that $|m|>|q|-1$. These correspond to  
the  regular central modes we  have discussed in the previous section.  
The discrete energy spectrum appears in the same conditions  
($\alpha>0$ and $\beta<0$) but now the quantization rules are
\begin{equation}
a_{1}=-n_{1}\,,\quad a_{2}=-n_{2}\,, \quad n_{1},\,n_{2}=0,1,2,...\,.
\end{equation}
Hereby we recover the  quantization rule (\ref{(abn)}) with the 
main quantum number of the regular modes $n=n_{1}+n_{2}+|m|+1$
and the same formula of the energy levels (\ref{(een)}). Moreover, we find 
that
\begin{equation}
\kappa=\sqrt{|\beta|}\,(n_{2}-n_{1}-q)\,.
\end{equation}
Thus the regular discrete axial modes  are labeled by the quantum numbers 
$n_{1},\,n_{2},\,m$ and $q$. Notice that $\alpha$ and $\beta$ depend only 
on $q$ and $n$ as it results from Eqs.(\ref{(ab)}) and (\ref{(een)}).

The irregular discrete modes can be derived for $\mu<0$ in the same 
manner like in the case of the central modes, by looking for the conditions 
in which the wave functions (\ref{(fh)}) remain square integrable on the 
domains $\xi,\eta>|\mu|$ for $s_{1},s_{2}<0$. We observe that this happens 
only when  $a_{1}=-n_{1},\, 
a_{2}=-n_{2},\, s_{1}=-(|m|+q)/2$ and  $s_{2}=-(|m|-q)/2$ 
such that $n_{1}<|m|+q-1$ and $n_{2}<|m|-q-1$. Hereby it results that 
the main quantum number of these  modes, $n=n_{1}+n_{2}-|m|+1$, accomplishes 
the selection rules $|q|<n<|m|-1$ and 
$\kappa=\sqrt{|\beta|}\,(n_{2}-n_{1}+q)$.  

\newpage
\section{Concluding remarks}
\

The study of the Schr\" odinger equation in Euclidean Taub-NUT 
space is well motivated. The Killing tensors of the Taub-NUT 
geometry imply the existence of extra conserved quantities, quadratic 
in four-velocities. The Taub-NUT case is analogous to the Coulomb 
problem where an extra degeneracy is present and there is the 
possibility of separating variables in two different coordinate systems.
This analogy is not perfect since in the Taub-NUT geometry the selection 
rules of the angular quantum numbers depend on $q$ and the boundary 
conditions  allow irregular modes. However, this  
has nothing surprising in general relativity where we know already 
the irregular modes of spinless \cite{AIS} or spin-half \cite{COTA} 
particles in anti-de Sitter space-times. 

In the last time, Iwai and Katayama \cite{IK1, IK2, IK3, YM} extended 
the Taub-NUT metric so that it still admits a Kepler-type symmetry. 
This class of metrics, of course, includes the original Taub-NUT 
metric. In general the Killing tensors involved in the Runge-Lenz 
vector cannot be expressed as a contracted product of Killing-Yano 
tensors. The only exception is the original Taub-NUT space \cite{MV}.
It will be interested to investigate the Schr\" odinger equation in the 
more complicated case of the generalized Taub-NUT spaces. The extension 
of the results for this class of space-times will be discussed elsewhere 
\cite{CV}.

\appendix

\section{$SO(3)\otimes U(1)$ harmonics}
\

The $SO(3)\otimes U(1)$  harmonics defined by 
Eqs.(\ref{(lp)})-(\ref{(spy)}) can not be expressed in 
terms of usual spherical harmonics. These are new harmonics that must be 
separately studied by solving the system of eigenvalue equations. We start 
with  
\begin{equation}\label{(Y)}
Y_{l,m}^{q}(\theta,\phi,\chi)=\frac{1}{4\pi}\, \Theta_{l,m}^{q}(cos\theta)
e^{im\phi}e^{iq\chi}
\end{equation}
since then Eqs.(\ref{(l3)}) and (\ref{(l0)}) are satisfied. It remains to 
solve only Eq.(\ref{(lp)}) and  calculate the normalization factor of 
$\Theta_{l,m}^{q}$   according to the condition
\begin{equation}\label{(np)}
\int_{-1}^{1}d(\cos\theta)\left|\Theta_{l,m}^{q}(\cos\theta)\right|^{2}=2\,.
\end{equation}  
To this end, we introduce the new variable $z=\sin^{2}\theta/2$
in Eq.(\ref{(lp)}) where ${\vect{L}\,}^{2}$ is given 
by (\ref{(lpa)}). Thus we obtain 
\begin{eqnarray}\label{(ecT)}
&&\left[z(1-z)\frac{d^2}{dz^2}+(1-2z)\frac{d}{dz}+ l(l+1)\right.\\
&&~~~~~~~~~~~~~~~~~~~~~~~~\left.-\frac{m^{2}+q^{2}-2(1-2z)mq}{4z(1-z)}\right]
\Theta_{l,m}^{q}(z)=0 \,.\nonumber
\end{eqnarray}
This equation has solutions of the form 
\begin{equation}\label{(Pz)}
\Theta_{l,m}^{q}(z)\sim z^{p}(1-z)^{k}\,F(p+k-l,\,p+k+l+1,\,2p+1,\, z)\,, 
\end{equation}
where the Gauss hypergeometric functions, $F$, depend on the real 
parameters $p$ and $k$ which satisfy $p^{2}=(m-q)^2/4$ and $k^{2}=(m+q)^2/4$.  

First of all we observe that the solutions (\ref{(Pz)}) are square integrable 
only when $F$ are polynomials selected by a quantization condition since 
otherwise $F$ becomes strongly divergent for $z\to 1$. This means that 
$l-p-k$ must be a non-negative integer number. If we replace the functions 
$F$ of (\ref{(Pz)})  by Jacobi polynomials \cite{AS}, we observe that the 
solutions of Eq.(\ref{(ecT)}) remain square integrable  if  $2p>-1$ and 
$2k>-1$. Like in the case of the usual spherical harmonics, the good choice 
is  $p=(|m|-q)/2$ and $k=(|m|+q)/2$
where 
\begin{equation}\label{(qml)}
|q|-1<|m|\le l\,.
\end{equation} 
Then by using the normalization condition  (\ref{(np)}) we find  the final 
result 
\begin{eqnarray}
\Theta_{l,m}^{q}(\cos\theta)&=&\frac{\sqrt{2l+1}}{2^{|m|}}\left[\frac{
(l-|m|)!\,(l+|m|)!}{
\Gamma(l-q+1)\Gamma(l+q+1)}\right]^{\frac{1}{2}}\label{(fin)}\label{(Tet)}\\
&&\times \left(1-\cos\theta\right)^{\frac{|m|-q}{2}}
\left(1+\cos\theta\right)^{\frac{|m|+q}{2}}\,
P_{l-|m|}^{(|m|-q,\,|m|+q)}(\cos\theta)\,.\nonumber
\end{eqnarray}
For $m=|m|$ the $SO(3)\otimes U(1)$ harmonics are given by (\ref{(Y)}) 
and (\ref{(fin)}) while for $m<0$ we have to use the obvious formula
\begin{equation}
Y_{l,-m}^{q}=(-1)^{m}\left(Y_{l,m}^{-q}\right)^{*}\,.
\end{equation} 
When the boundary conditions allow half-integer quantum numbers $l$ and $m$
then we say that the functions defined by Eqs.(\ref{(Y)}) and (\ref{(fin)}) 
 (up to a suitable factor) represent $SU(2)\otimes U(1)$ harmonics.

Thus we have obtained a non-trivial generalization of the  spherical 
harmonics of the same kind as the spin-weighted spherical 
harmonics \cite{NP}. Indeed, if $l$, $m$ and $q=m'$ are either integer or 
half-integer numbers then we have  
\begin{equation}
Y_{l,m}^{m'}(\theta,\phi,\chi)=\frac{\sqrt{2l+1}}{4\pi}D^{l}_{m,m'}
(\phi,\theta, \chi) 
\end{equation}
where $D_{m,m'}^{l}$ are the matrix elements of the irreducible representation 
of weight $l$ of the $SU(2)$ group corresponding to the rotation of Euler 
angles $(\phi, \theta, \chi)$. What is new in the case of our harmonics is 
that these are defined for any real number $q$.

\end{document}